\documentclass[12pt,a4paper]{article}
\usepackage[latin1]{inputenc}
\usepackage[T1]{fontenc}
\usepackage{url}

\usepackage{amsmath}
\usepackage{amsfonts}
\usepackage{amssymb}
\usepackage{graphicx}
\usepackage{epsfig}
\usepackage{cite}
\newcommand{\datm}{\Delta m_{atm}^2}
\newcommand{\dsol}{\Delta m_{sol}^2}
\makeatletter
\renewcommand*\env@matrix[1][*\c@MaxMatrixCols c]{%
  \hskip -\arraycolsep
  \let\@ifnextchar\new@ifnextchar
  \array{#1}}
\makeatother

\begin{document}
\title{\bf \large On the importance of the 1-loop finite\\ corrections to seesaw
neutrino masses}
\author{D. Aristizabal Sierra\footnote{\tt daristizabal@ulg.ac.be}\\
\small \sl IFPA, Dep. AGO, Universite de Liege, Bat B5\\ \small \sl Sart
Tilman B-4000 Liege 1, Belgium\\
\and
Carlos E. Yaguna\footnote{\tt carlos.yaguna@physik.uni-wuerzburg.de}\\
\small \sl Institute f\"ur Theoretische Physik und Astrophysik,\\
\small \sl Universit\"at W\"urzburg, 97074 W\"urzburg, Germany
}

\date{}
%\pacs{}
%\keywords{}
\maketitle
\begin{abstract}
In the standard seesaw mechanism, finite corrections to the neutrino mass matrix
arise from 1-loop self-energy diagrams mediated by a heavy neutrino. We study in detail these
corrections and demonstrate that they can be very significant, exceeding in several cases the tree-level
result. We consider the normal and inverted hierarchy spectra
for light neutrinos and compute the finite corrections to the different
elements of the neutrino mass matrix. Special attention is paid to their dependence with the
parameters of the seesaw model. Among the cases in which the corrections  can be large, we identify the fine-tuned models considered previously in the
literature, where a strong cancellation between the different parameters is
required to achieve compatibility with the experimental data. As a particular example, we also analyze how these corrections modify the tribimaximal mixing pattern and find that the deviations may be sizable, in particular for $\theta_{13}$. Finally, we emphasize that due to their large size, the finite corrections to neutrino masses have to be taken into account if one wants to properly scan the parameter space of seesaw models.
\end{abstract}
\newpage
\section{Introduction}
\label{sec:intro}
Neutrino oscillation experiments have firmly established that neutrino
have tiny but non-zero masses and that the mixing in the leptonic
sector is in sharp contrast with the small mixing that characterizes
the quark sector \cite{Schwetz:2008er,GonzalezGarcia:2010er}. From a
theoretical perspective the smallness of neutrino masses can be well
understood within the seesaw model
\cite{Minkowski:1977sc,Yanagida:1979as,Mohapatra:1979ia,GellMann:1980vs,Schechter:1980gr},
in which the fermion sector of the Standard Model is extended by
adding new electroweak fermionic singlets (standard seesaw
model). In this framework,  light neutrino masses are generated via mixing with the singlet states and their smallness can naturally be explained if the singlets masses are very large.

The determination of the regions of parameter space consistent with
low energy neutrino observables in the seesaw model typically relies
on parametrizations of the neutrino Yukawa couplings
\cite{Casas:2001sr,Davidson:2001zk}. Once the Yukawas are properly
parametrized, such regions are found by doing
numerical scans in which the neutrino experimental data is used as an
input. This procedure is always based on the tree-level light neutrino
mass matrix and fails  if in
some regions of the parameter space the one-loop corrections to the tree-level
mass matrix turn out to have sizable values.

The one-loop corrections to the seesaw light neutrino mass matrix  were first discussed in
\cite{Grimus:1989pu} in a general setup with an arbitrary number of singlets, lepton doublets and Higgs doublets. They were
later analyzed in ref. \cite{Pilaftsis:1991ug} in a particular
realization in which, due to a particular Yukawa mass matrix, light
neutrino masses vanish at tree-level and are entirely generated by the
one-loop corrections.  Subsequently, the renormalization of general theories
with Dirac and/or Majorana neutrinos was carried out in
\cite{Kniehl:1996bd} and additional studies were done in references
\cite{Grimus:1999wm,Grimus:2001ex,Pilaftsis:2002nc,Grimus:2002ux,Grimus:2002nk}.

Loop corrections are of two types: renormalizable and intrinsically
finite. The renormalizable pieces consist of corrections to the tree
level parameters already present in the seesaw Lagrangian, and are
suppressed with respect to the tree level piece by extra Yukawa
couplings and by the loop factor 1/16$\pi^2$. The finite parts instead
are corrections to the vanishing elements of the tree-level mass matrix for the neutral fermions, and are only suppressed by the loop
factor. Thus, they are  potentially large.

The aim of this paper is to quantify the impact that the finite one-loop
corrections might have on the effective light neutrino mass
matrix. We consider the most general standard seesaw model and
numerically analyze the importance of these corrections for the
different mass matrix elements as well as for the neutrino mass eigenvalues and mixing angles, differentiating in our discussion
between the normal and the inverted light neutrino mass spectrum. We
will show that the corrections can range over several orders of
magnitude, depending on whether one relies or not on models where
consistency with the measured neutrino masses and mixing angles
requires strong cancellations in the tree-level neutrino mass
matrix. Indeed, as we will discuss, once one-loop corrections are
taken into account these models are barely reconcilable with data.
Barring these cases, we will prove that the finite corrections are usually of order $~20-40\%$
but they may  exceed the tree-level value for certain entries that can be
strongly suppressed. In order to
make reliable predictions in the seesaw, therefore, the finite one-loop corrections
should be included. 

The rest of the paper is organized as follows: in section
\ref{sec:seesaw-tree-level} we define our notation and briefly
describe the seesaw model at tree-level, including its  standard
parametrization. We discuss the finite 1-loop
corrections in
section~\ref{sec:one-loop-corrections}. In sections \ref{sec:nor-spectrum} and
\ref{sec:inv-spectrum}  our main results are presented --the calculation of the finite corrections  for the normal and inverted light neutrino mass spectrum. Then, we consider the particular case of tribimaximal mixing in section \ref{sec:tribi} and we determined how such mixing pattern is modified by the 1-loop finite corrections. A brief  discussion of our results and some comments on their possible phenomenological
implications is given in section \ref{sec:discussion}. In section \ref{sec:conclusions} we draw our conclusions
and summarize our findings. For completeness, in the appendix we
present the details of the calculation of the finite one-loop
corrections.
\section{The standard seesaw model at tree level}
\label{sec:seesaw-tree-level}
In the standard seesaw model, three fermionic electroweak singlets
$N_{R_i}$ are added to the Standard Model. In
the basis in which the matrix of charged lepton Yukawa couplings and the
singlet mass matrix are diagonal the Lagrangian accounting
for the new interactions can be written as
\begin{equation}
  \label{eq:seesaw-Lag}
  -{\cal L}=-i\bar N_{R_i}\,{\partial\!\!\!\!\diagup} N_{R_i}
  %+ \phi^\dagger\bar\ell_{R_i}Y_i \ell_{L_i}
  + \tilde\phi^\dagger\bar N_{R_i}\lambda_{ij}\ell_{Lj}
  + \frac{1}{2}\bar N_{R_i} C M_{R_i} \bar N_{R}^T
  + \mbox{h.c.}
\end{equation}
where $\phi^T=(\phi^+ \phi^0)$ is the Higgs electroweak doublet,
$\ell_L$ are the leptonic $SU(2)$ doublets, $C$
is the charge conjugation operator and
$\mathbf{\lambda}$ is a Yukawa matrix in flavor space. In the left-handed
chiral basis $\mathbf{n_L}^T=(\mathbf{\nu_L}, (\mathbf{N_R})^C)^T$, and once electroweak
symmetry breaking is taken into account, the neutral fermion mass terms can be written
as
\begin{equation}
  \label{eq:seesaw-Lag-rewritten}
  -{\cal L}_{F^0}= \frac{1}{2}\mathbf{n}_{\mathbf{L}}^T C
  \mathbf{\cal M}\,\mathbf{n_L}
  + \mbox{h.c.}
\end{equation}
where $\mathbf{\cal M}$, the $6\times 6$ neutral fermion mass matrix, is given by
\begin{equation}
  \label{eq:seesaw-mass-matrix}
  \mathbf{\cal M}=
  \begin{pmatrix}[cc]
    \mathbf{0~} & \mathbf{M}_{\mathbf{D}}^T\\
    \mathbf{M_D} & \mathbf{\hat M_R}
  \end{pmatrix}\,,
\end{equation}
with $\mathbf{M_D}=v\mathbf{\lambda}$ ($v=\sqrt{2} M_W/g\simeq 174$
GeV). The mass spectrum is obtained by rotating the fields to the mass
eigenstate basis, denoted by $\chi_i$, via the unitary matrix $\mathbf{U}$:
\begin{equation}
\mathbf{\chi_L}=\mathbf{U}^\dagger \mathbf{n_L}\,.
\end{equation}
In this basis, the Lagrangian mass terms, equation (\ref{eq:seesaw-Lag-rewritten}), become
\begin{equation}
  \label{eq:mass-eig-Lag}
  -{\cal L}_{F^0}=\frac{1}{2}\overline{\mathbf{\chi}}\,
  \mathbf{\hat{\cal M}}P_L\,
  \mathbf{\chi} + \text{h.c.}
\end{equation}
where
\begin{equation}
\label{eq:diag-matrix}
 \mathbf{U}^T\,\mathbf{\cal M}\,\mathbf{U}=
\mathbf{\hat{\cal M}}=\mbox{diag}(m_{\chi_1},\dots m_{\chi_6})
\end{equation}
and $\chi_i$ are the physical Majorana neutrino fields. Note that by decomposing
the matrix $\mathbf{U}$ as \cite{Grimus:1989pu,Grimus:2002nk,Grimus:2002ux}
\begin{equation}
  \label{eq:matrix-U}
  \mathbf{U}=
  \begin{pmatrix}[l]
    \mathbf{U_L}\\
    \mathbf{U}_\mathbf{R}^*
  \end{pmatrix}
\end{equation}
the $\nu_{L_i}$ and $N_{R_i}$ states can be expressed as
\begin{equation}
  \label{eq:nu-NR}
  \nu_{L_i}=U_{L_ij}P_L \chi_j\,,\quad N_{R_i}=U_{R_{ij}}P_R\chi_j\,.
\end{equation}
In the seesaw limit ($\mathbf{M_D}\ll \mathbf{M_R}$),
the diagonalization of the mass matrix (\ref{eq:seesaw-mass-matrix})
gives rise to a split spectrum consisting of  three heavy states with masses $M_{R_i}$ and three light states with an effective mass matrix,      $\mathbf{m}_{\boldsymbol{\nu}}^{\mbox{\tiny(tree)}}$,  given by
\begin{equation}
  \label{eq:seesaw-formula-tree-level}
\mathbf{m}_{\boldsymbol{\nu}}^{\mbox{\tiny(tree)}}=
-\mathbf{M}_{\mathbf{D}}^T\,\mathbf{\hat M_R}^{-1}\,\mathbf{M_D}\,.
\end{equation}
The light neutrino mass spectrum, mixing angles and CP violating
phases --the so-called low-energy observables-- are obtained from this matrix after
diagonalization:
\begin{equation}
  \label{eq:diagonalization}
  \mathbf{U}_\ell^T\mathbf{m}_{\boldsymbol{\nu}}^{\mbox{\tiny (tree)}}\mathbf{U}_\ell=
  \mathbf{\hat m_\nu}\,,
\end{equation}
where $\mathbf{U_\ell}$ is the leptonic mixing matrix parametrized
according to
\begin{equation}
  \label{eq:leptonic-mix-matrix}
  \mathbf{U_\ell}=\mathbf{U_\ell}(\theta_{23})\mathbf{U_\ell}(\theta_{13},\delta)
  \mathbf{U_\ell}(\theta_{12})\times\mbox{diag}(e^{-i\varphi_1},e^{-i\varphi_2},1)
\end{equation}
with $\delta, \varphi_{1,2}$ being respectively the Dirac and Majorana CP violating
phases and $U_\ell(\theta)$ rotation matrices.

The determination of the seesaw parameters compatible with neutrino
experimental data relies on parametrizations of the Yukawa couplings or, equivalently, of the Dirac neutrino mass matrix,
$\mathbf{M_D}=v\,\mathbf{\lambda}$. In the numerical analysis of the finite
one-loop corrections, we have used the most common parametrization of the seesaw, the Casas-Ibarra parametrization
\cite{Casas:2001sr}.  In this  parametrization, the most general
$\mathbf{M_D}$ compatible with eq. (\ref{eq:seesaw-formula-tree-level})
is given by
\begin{equation}
  \label{eq:defR}
  \mathbf{M_D}=i\,\mathbf{\hat M_R}^{1/2}\,\mathbf{R}\,
  \mathbf{\hat m_\nu}^{1/2}\,\mathbf{U_\ell}^\dagger\;,
\end{equation}
where $\mathbf{R}$ is any orthogonal matrix. This matrix can be written as a
rotation matrix determined by three complex angles.

The neutrino mass eigenvalues and the mixing matrix entering into this
equation are strongly constrained by experimental data whereas the
masses of the singlet neutrinos and the matrix $\mathbf{R}$ are entirely
free parameters of the seesaw model. In the numerical treatment of our results (sections \ref{sec:nor-spectrum} and \ref{sec:inv-spectrum})
we also impose the perturbativity condition suggested recently in
\cite{Casas:2010wm}:
\begin{equation}
 \mathrm{Tr}\left[\mathbf{\lambda}^\dagger \mathbf{\lambda}\right]\leq 3.
\end{equation}
Now that we have reviewed the seesaw mechanism at tree level, let us take a look at the 1-loop corrections to neutrino masses.

\section{Finite one loop corrections to the neutral\\fermion mass matrix}
\label{sec:one-loop-corrections}
In the standard seesaw, the  one-loop corrections to the $\nu-N$ mass matrix  are determined by the neutrino interactions with the $Z$ boson, the neutral Goldstone bosons ($G^0$), and the Higgs boson
($h^0$) --see appendix
\ref{sec:self-energies}. All together, in
addition to the correction involving the standard model leptonic
charged current, they define the one-loop two point function
$-i\mathbf{\Sigma}(p)$ \cite{Grimus:2002nk}.

Once the one-loop corrections are taken into account
the neutral fermion mass matrix is given by
\begin{equation}
  \label{eq:neutral-fermion-mm-treeloop}
  \mathbf{\cal M}=\mathbf{\cal M}^{{\mbox{\tiny(tree)}}}
  +\mathbf{\cal M}^{{\mbox{\tiny(1-loop)}}}\,,
\end{equation}
where the 1-loop contribution can be decomposed as
\begin{equation}
  \label{eq:one-loop-mass-matrix}
  \mathbf{{\cal M}}^{{\mbox{\tiny(1-loop)}}}=
  \begin{pmatrix}
    \mathbf{\delta M_L} & \mathbf{\delta M}_{\mathbf{D}}^T\\
    \mathbf{\delta M_D}        & \mathbf{\delta M_R}
  \end{pmatrix}\,.
\end{equation}
Notice that the $\mathbf{0}_{3\times 3}$ matrix
appearing at tree-level is replaced by the contribution
$\mathbf{\delta M_L}$, which among all the sub-matrices in
$\mathbf{{\cal M}}^{\mbox{\tiny (1-loop)}}$ is the dominant one
\cite{Grimus:2002nk}.

Neglecting the subdominant pieces in $\mathbf{{\cal M}}^{\mbox{\tiny (1-loop)}}$ and
after block diagonalization of the neutral fermion mass matrix, the effective
light neutrino mass matrix, up to one-loop order, can be written as
\begin{equation}
  \label{eq:eff-nmm-1loop}
    \mathbf{m_\nu}=
    \mathbf{m}_{\nu}^{\mbox{\tiny(tree)}} + \mathbf{m}_{\boldsymbol{\nu}}^{\mbox{\tiny(1-loop)}}=
    -\mathbf{M}_{\mathbf{D}}^T\,\mathbf{\hat M_R}^{-1}\,\mathbf{M_D} + \mathbf{\delta M_L}\,.
\end{equation}
The sub-matrix $\mathbf{\delta M_L}$ and all the other sub-matrices entering in
$\mathbf{\cal M}^{{\mbox{\tiny(1-loop)}}}$  are entirely determined by the self-energy
functions $\mathbf{\Sigma_L^S}(p^2)$ (see appendix \ref{sec:self-energies}) via the diagonalization relation (\ref{eq:diag-matrix}):
\begin{equation}
  \label{eq:loop-parameters}
  \mathbf{\cal M}^{{\mbox{\tiny(1-loop)}}}=\mathbf{U}^*\mathbf{\Sigma_L^S}(p^2)\mathbf{U}^\dagger\,.
\end{equation}
Accordingly, the finite contribution is given by
\begin{equation}
  \label{eq:finite-contribution}
  \mathbf{\delta M_L}=\mathbf{U_L}^*\mathbf{\Sigma^S_L}(p^2)\mathbf{U_L}^\dagger
  =\mathbf{U_L}^*\mathbf{\Sigma^S_L}(0)\mathbf{U_L}^\dagger\,,
\end{equation}
where we have used the fact that $\mathbf{\Sigma^S_L}$ can be evaluated
at zero external momentum \cite{Grimus:2002nk}.
The self-energy functions $\mathbf{\Sigma^S_L}(0)$ are determined by
three Feynman self-energy diagrams involving the $Z$, the neutral
Goldstone boson $G^0$ and the Higgs boson $h^0$. Each diagram contains
a divergent piece but when summing up the three contributions the
result turns out to be finite, as it has to be since there are no
counterterms that would allow to absorb a possible  divergence
(see appendix \ref{sec:self-energies} for more details). The final expression
for the finite one-loop correction is given by\cite{Grimus:2002nk}
\begin{equation}
   \label{eq:deltaML}
 \mathbf{\delta M_L}=\mathbf{M}_{\mathbf{D}}^T \mathbf{\hat M_R}^{-1}
  \left\{\frac{g^2}{64 \pi^2 M_W^2}
  \left[
    m_h^2\ln\left(\frac{\mathbf{\hat M_R}^2}{m^2_h}\right)
    +
    3 M_Z^2\ln\left(\frac{\mathbf{\hat M_R}^2}{M^2_Z}\right)
  \right]\right\}\mathbf{M_D}\,.
\end{equation}
Notice that this correction is not suppressed, with respect to tree-level result, by additional factors of $M_D/M_R$. Thus, it is expected to be smaller than the tree-level mass term solely by a factor of order $(16\pi^2)^{-1}\ln(M_R/M_Z)$.  

In spite of the similar structure of the  1-loop  correction and the tree-level result, they are not proportional to each other unless the heavy neutrinos are degenerate --$\mathbf{M_R}\propto \mathbf{I}$. Hence, one could in principle have that $\mathbf{m}_{\nu}^{\mbox{\tiny (tree)}}=0$ and that neutrino masses arise entirely from 1-loop effects, as proposed in \cite{Pilaftsis:1991ug}.  Such models, however, are rather contrived and will not be discussed in the following. We are interested, instead, in the generic modifications to the neutrino mass matrix induced by the 1-loop corrections.

To evaluate these corrections, we first find sets of $\mathbf{M_D}$ and $\mathbf{\hat M_R}$ compatible with the experimental data at tree-level --using equation(\ref{eq:defR})-- and then use them to evaluate $\mathbf{\delta M_L}$\footnote{Alternatively, one could choose the seesaw parameters so that at 1-loop they are compatible  with the experimental data. Both procedures give rise to the same effects.}. Specifically, we generate the diagonal matrix of light neutrino masses (according to the desired spectra: normal or inverted) and the mixing matrix $\mathbf{U}_\ell$ such that they are compatible with neutrino data. For simplicity, the phases in $\mathbf{U_\ell}$ were assumed to vanish. Then, we randomly generate the three masses of the heavy states (in the range $1$ TeV to $10^{12}$ GeV) and the elements of the orthogonal matrix $\mathbf{R}$. From equation (\ref{eq:defR}), we can then obtain $\mathbf{M_D}$, which together with the generated $\mathbf{M_R}$ allows us to evaluate $\mathbf{\delta M_L}$\footnote{In our analysis we fix $m_{h^0}=150$ GeV.}. The size of the corrections is then determined by the ratio between the contributions up to 1-loop order and the tree-level result for the different elements of the neutrino mass matrix,
$(\mathbf{m}_\nu^{\mbox{\tiny (tree)}} + \mathbf{m}_{\nu}^{\mbox{\tiny (1-loop)}})/\mathbf{m}_\nu^{\mbox{\tiny(tree)}}$.

In the next two sections our main results are presented: we compute the corrections to the neutrino mass matrix in the seesaw model for the two different kinds of light neutrino spectra, with normal and inverted hierarchy.

\section{Corrections for the Normal Hierarchy spectrum}
\label{sec:nor-spectrum}
If the spectrum of light neutrinos has a normal hierarchy ($m_{\nu_3}=\sqrt{ \datm}$,
 $m_{\nu_2}=\sqrt{\dsol}$, $m_{\nu_1}\ll m_{\nu_2},m_{\nu_3}$), the elements of the neutrino mass matrix take
values within the following ranges:
\begin{equation}
\mathbf{m}_\nu^{\mbox{\tiny{exp}}}=\left(\begin{array}{rrr}
(2.5,5.5) \times 10^{-12} & (2.3,9.8)\times 10^{-12}
& (-3.3,4.9)\times 10^{-12}\\
-\hspace{1cm} & (2.0,3.4)\times 10^{-11} & (1.9,2.3)\times 10^{-11}\\
-\hspace{1cm} & -\hspace{1cm}& (2.1,3.4)\times 10^{-11}
\end{array}
\right)
\,\mbox{GeV}
\label{eq:numasses}
\end{equation}
as  the oscillation parameters vary within their $2$-$\sigma$
experimentally allowed intervals \cite{GonzalezGarcia:2010er,Schwetz:2008er}.  Since the
matrix is symmetric, we only show the six independent matrix elements. Notice, in
particular, that the element ($1,3$) is the only one that can vanish in this case. It can be easily checked that this can happen if $\theta_{13}$ is between $4^\circ$ and $6^\circ$. All other elements vary within a relatively small range --not so small for (1,2)-- between $10^{-11}$ and $10^{-12}$ GeV. Since the corrections to the neutrino mass matrix are
not proportional to the matrix element itself, the correction to the element
($1,3$)  could easily exceed its tree-level value.

With the aim of facilitating the study of these corrections and the understanding of their origin,
we will divide our analysis in two parts depending on what is assumed for the orthogonal matrix
$\mathbf{R}$. First it is taken to be  real and then
the most general case is considered, a complex matrix. The number of free parameters will therefore increase as we move from the first case to the second.
\subsection{$\mathbf{R}$ real}
\begin{figure}[tb]
\begin{center}
\includegraphics[scale=0.45]{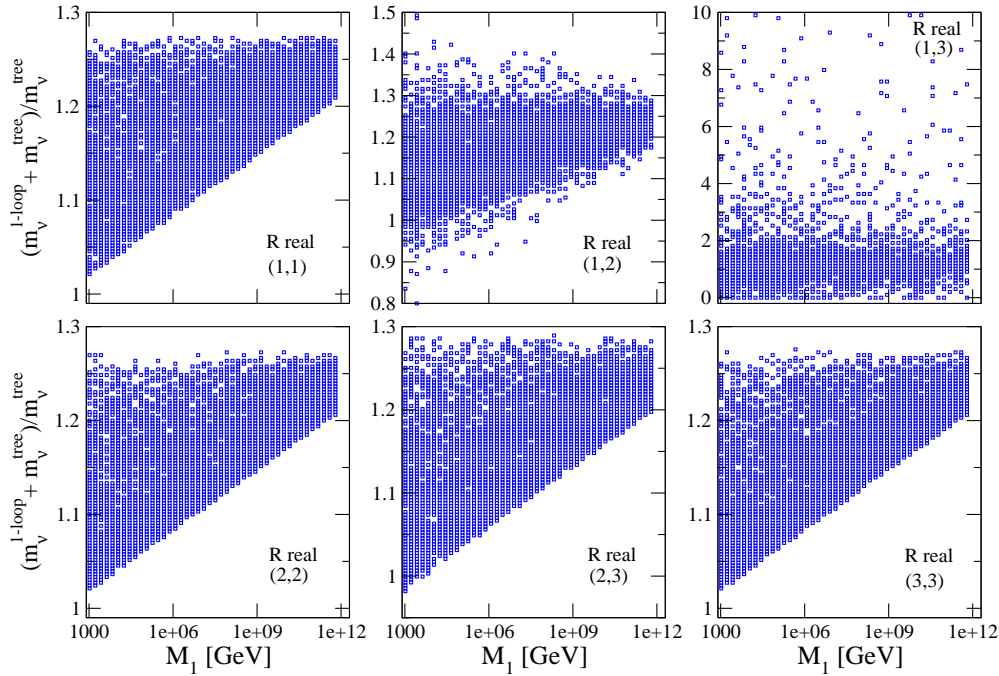}
\caption{\small \it The ratio between the 1-loop and the tree-level result for
the different elements of the neutrino mass matrix as a function of $M_1$.
It has been assumed that $\mathbf{R}$ is real and that light neutrinos have a NH
spectrum.
\label{fig:harrayRR}}
\end{center}
\end{figure}

If $\mathbf{R}$ is real, the parameters needed to evaluate the 1-loop correction to the neutrino mass matrix are the oscillation parameters, the three masses of the right handed
neutrinos ($M_i$), and  the three angles that
parametrize $\mathbf{R}$. To obtain the numerical results below, we vary the neutrino mixing angles and mass squared differences within their $2\sigma$ ranges, and  we randomly choose $M_i$ between $1$ TeV and $10^{12}$ GeV and the angles in $\mathbf{R}$ between  $0$ and $2\pi$.

The resulting corrections to the matrix elements are shown in figure \ref{fig:harrayRR} as a function of the mass of the lightest heavy neutrino, $M_1$. We see that they are similar for the elements ($1,1$), ($2,2$), ($2,3$), and ($3,3$), increasing with $M_1$ and reaching values up to order $30\%$. Those for the  element
($1,2$) are slightly different, reaching values  as large as $40\%$ or $50\%$ as well
as  $-20\%$.

The corrections to the element ($1,3$), on the other hand, can be quite  large for a
significant fraction of models. As anticipated, this result is due to the fact that the element ($1,3$) can be very small so it may receive a huge fractional correction. It must be noticed in that case, however, that a large value of $(\mathbf{m}_\nu^{\mbox{\tiny (tree)}} + \mathbf{m}_{\nu}^{\mbox{\tiny (1-loop)}})/\mathbf{m}_\nu^{\mbox{\tiny(tree)}}$ does not necessarily imply a significant deviation in the expected value of the neutrino observables --the mass eigenvalues and the mixing angles. For that reason it is important to study the effect of the corrections on both the  matrix elements and the predicted observables. We will do so in the next section, where we consider the most general case:  $\mathbf{R}$ complex. 

Notice then that even in the case $\mathbf{R}$ real, where no large parameters are introduced, the corrections to neutrino masses can be quite important. If $M_1\gtrsim 10^9$ GeV they are expected to be larger than about $15\%$ and they could easily reach $25\%$ or $30\%$.
\subsection{$\mathbf{R}$ complex}
\begin{figure}[t!]
\begin{center}
\includegraphics[scale=0.45]{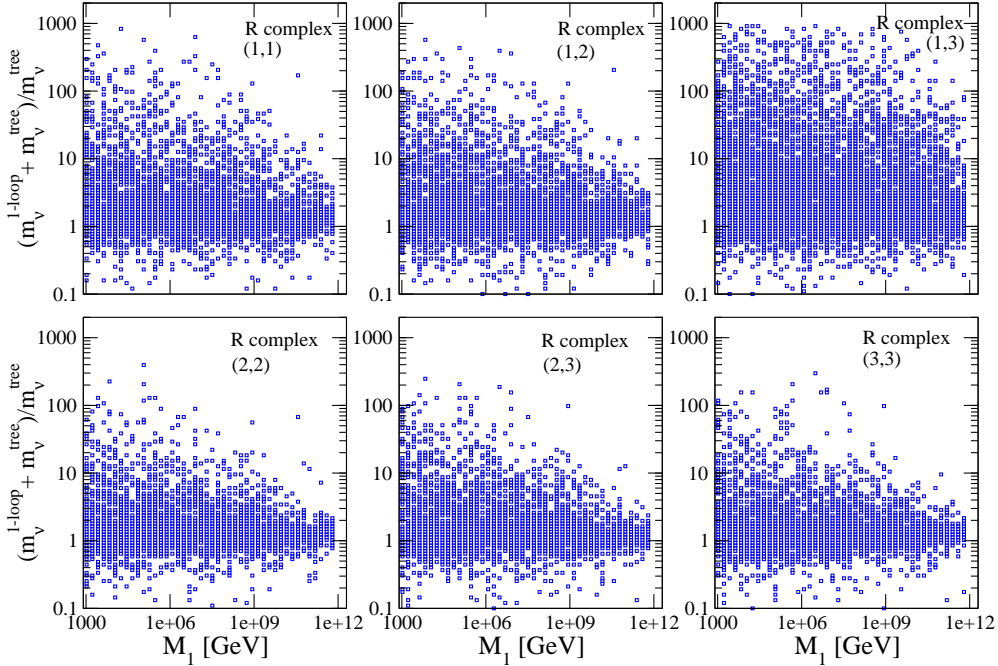}
\caption{\small \it The ratio between the 1-loop and the tree-level result for
the different elements of the neutrino mass matrix as a function of $M_1$.
It has been assumed that $\mathbf{R}$ is complex and that light neutrinos have a NH
spectrum.
\label{fig:harrayRC}}
\end{center}
\end{figure}

\begin{figure}[t!]
\begin{center}
\includegraphics[scale=0.45]{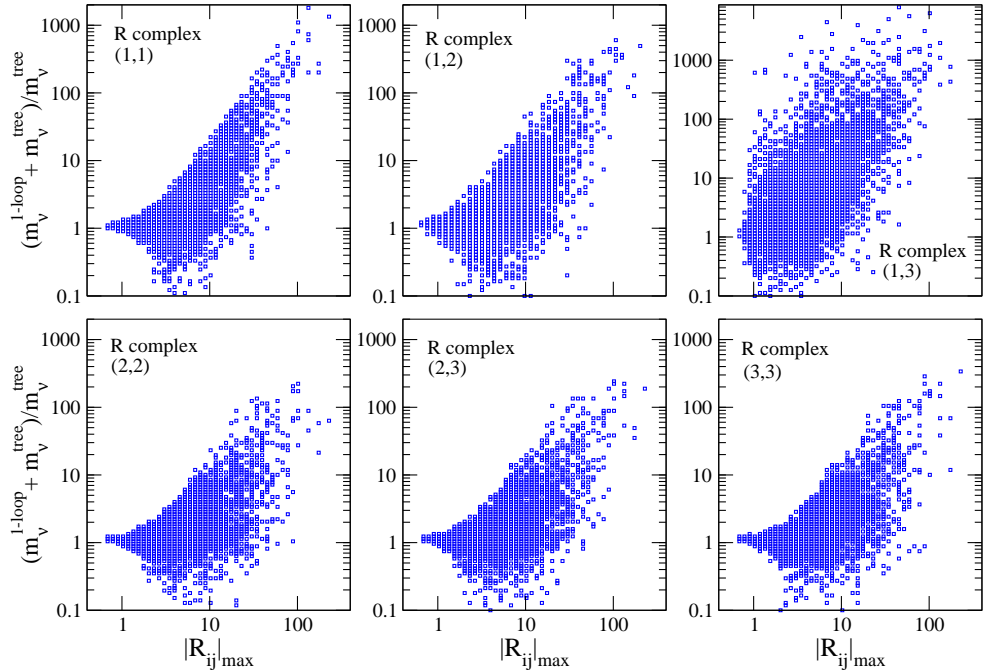}
\caption{\small \it The ratio between the 1-loop and the tree-level result for
the different elements of the neutrino mass matrix as a function of the
largest element of the $\mathbf{R}$ matrix.
It has been assumed that $\mathbf{R}$ is complex and that light neutrinos have a NH
spectrum.
\label{fig:harrayrmaxRC}}
\end{center}
\end{figure}

 This case is not only the most general one,
but it is also
well motivated by leptogenesis. In fact, in the case of unflavored leptogenesis, the phases in $\mathbf{R}$ are the ones
responsible for the CP-asymmetry in the decays of the heavy neutrinos and
ultimately for the generation of the baryon asymmetry. 
%in leptogenesis
%scenarios in which leptogenesis takes place at temperatures above $10^{12}$ GeV.

In this case, the three angles parametrizing $\mathbf{R}$ are complex numbers, with a
certain magnitude and a given phase. In our analysis, we allow these complex
angles to have  an arbitrary phase and we restrict their magnitude
to be smaller than $3$\footnote{One can relax this assumption to obtain even larger effects.}. Since $\cosh 3\sim \sinh 3\sim 10$, $\mathbf{R}$ can have
elements at most of order $10^3$. When the elements of  $\mathbf{R}$ are significantly larger than $1$, $|R_{ij}|\gg 1$, one obtains the so-called fine-tuned models. In them, strong cancellations between the different terms in equation (\ref{eq:seesaw-formula-tree-level}) are required to obtain compatibility with the experimental data.  Since the corrections to the
neutrino mass matrix depend on $\mathbf{R}$, the loop suppression in (\ref{eq:deltaML}) can be easily overcome by the large elements in $\mathbf{R}$, yielding a correction that is  significantly larger than the tree level result.

Figure \ref{fig:harrayRC} shows the corrections to the different matrix
elements for $\mathbf{R}$ complex. We see that, in fact, the 1-loop contribution can
exceed, for all the matrix elements, the three-level result by several orders of
magnitude. In that case, there is no doubt that the corrections will have a huge impact on the predicted neutrino mass eigenvalues and mixing angles.

\begin{figure}[t!]
\begin{center}
\includegraphics[scale=0.45]{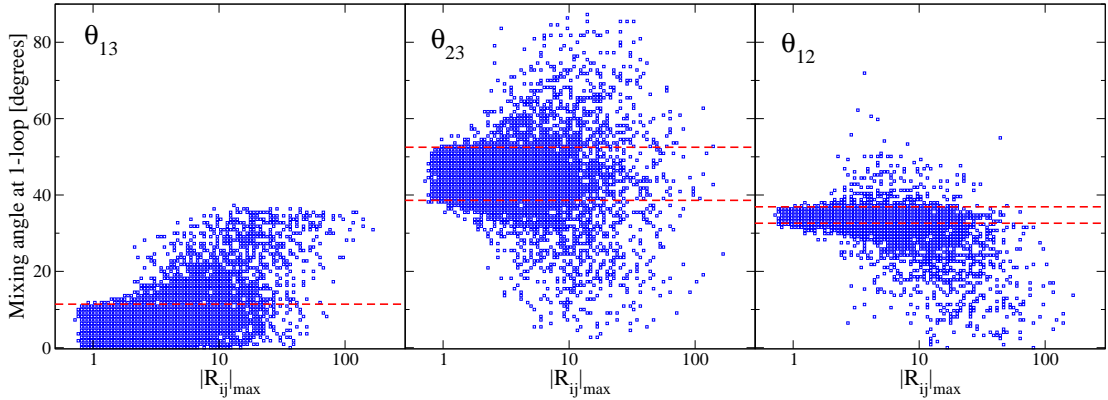}
\caption{\small \it The  neutrino mixing angles at 1-loop as a function of the
largest element of the $\mathbf{R}$ matrix.
It has been assumed that $\mathbf{R}$ is complex and that light neutrinos have a NH
spectrum. The region between the two dashed (red) lines is consistent with current experimental data at $2\sigma$. 
\label{fig:norHangles}}
\end{center}
\end{figure}

\begin{figure}[t!]
\begin{center}
\includegraphics[scale=0.45]{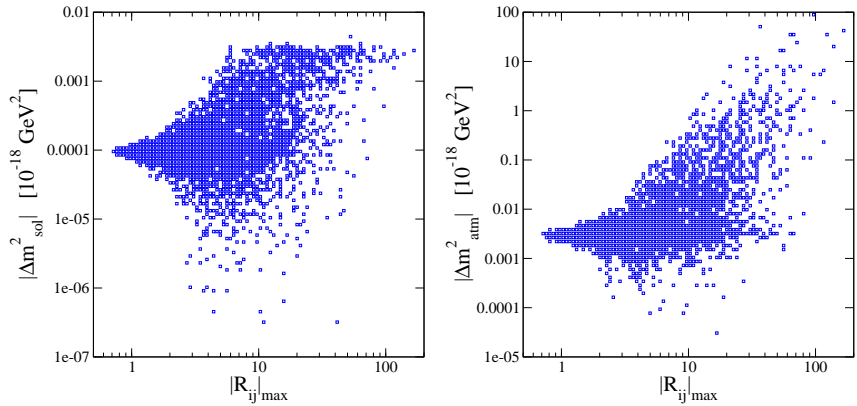}
\caption{\small \it The  neutrino mass squared differences at 1-loop as a function of the
largest element of the $\mathbf{R}$ matrix.
It has been assumed that $\mathbf{R}$ is complex and that light neutrinos have a NH
spectrum.
\label{fig:norHmasses}}
\end{center}
\end{figure}

It is indeed the large elements present in $\mathbf{R}$ that make possible a
1-loop correction much larger than the tree-level result. We illustrate
this fact in figure \ref{fig:harrayrmaxRC}, which shows the size of the
corrections as a function of the largest element of the $\mathbf{R}$ matrix, $\left|R_{ij}\right|_{\mbox{\tiny{max}}}$. Notice that
when this element is of order $1$ the corrections are usually small (those to the element ($1,3$) being the exception) but they increase
with it reaching two orders of magnitude or more for $\left|R_{ij}\right|_{\mbox{\tiny{max}}}$
around $100$. With such huge corrections, the agreement between the tree-level seesaw formula and the neutrino data assumed in the parametrization becomes meaningless. In fact, as illustrated in figures \ref{fig:norHangles} and  \ref{fig:norHmasses}, the oscillation parameters may deviate significantly from their observed values once the 1-loop corrections are taken into account.

Figure \ref{fig:norHangles} displays the 1-loop mixing angles, those obtained from the diagonalization of the neutrino mass matrix at 1-loop, as a function of $\left|R_{ij}\right|_{\mbox{\tiny{max}}}$. The region consistent with the experimental data at $2\sigma$ is the area between the two dashed lines. Notice that all the angles, which were chosen to be consistent with the data at tree level, can at 1-loop become much larger than allowed by present observations. 

Similarly, we see in figure \ref{fig:norHmasses} that the neutrino mass squared  differences at 1-loop can vary over several orders of magnitude. A fact that is in clear contradiction with current experiments. 

As we have seen, for the neutrino spectrum with normal hierarchy the corrections to the neutrino mass matrix can be quite important. The matrix element ($1,3$), in particular, can receive very large fractional corrections independently of $\mathbf{R}$ if $m_{\nu13}$ is suppressed. In addition, all matrix elements as well as the neutrino mass eigenvalues and mixing angles are expected to receive significant corrections  when the elements of the $\mathbf{R}$ matrix are  larger than one. In such case, the inclusion of the 1-loop corrections is mandatory.

Next, we analyze the importance of these corrections for the neutrino spectrum with inverted hierarchy.

\section{Corrections for the Inverted Hierarchy spectrum}
\label{sec:inv-spectrum}
If the spectrum of light neutrinos has an inverted hierarchy (IH),
$m_{\nu_2}=\sqrt{\left|\datm\right|}$,
 $m_{\nu_1}=\sqrt{\left|\datm\right|-\dsol}$, $m_{\nu_3}\ll m_{\nu_2},m_{\nu_1}$, the elements of the neutrino
mass matrix take values within the following ranges:
\begin{equation}
\mathbf{m_\nu}^{\mbox{\tiny{exp}}}=\left(\begin{array}{rrr}
(4.5,5.1) \times 10^{-11} & (-8.5,1.7)\times 10^{-12}
& (-8.5,1.3)\times 10^{-12}\\
-\hspace{1cm} & (1.7,3.3)\times 10^{-11} & (-2.5,-2.0)\times 10^{-11}\\
-\hspace{1cm} & -\hspace{1cm}& (1.8,3.4)\times 10^{-11}

\end{array}
\right)
\,\mbox{GeV}
\end{equation}
as  the oscillation parameters vary within their $2$-$\sigma$
experimentally allowed intervals \cite{GonzalezGarcia:2010er,Schwetz:2008er}. Notice  that in
this case the elements ($1,2$)  and ($1,3$) are both allowed to vanish, an event that can happen if $\theta_{13}$ is smaller than about $2^\circ$. We expect, therefore, large fractional corrections to the
entries ($1,2$) and ($1,3$) of the neutrino mass matrix independently of
$\mathbf{R}$. As before, we will divide the analysis of the 1-loop corrections into two parts: $\mathbf{R}$ real and $\mathbf{R}$ complex.
\subsection{$\mathbf{R}$ real}
\begin{figure}[tb]
\begin{center}
\includegraphics[scale=0.45]{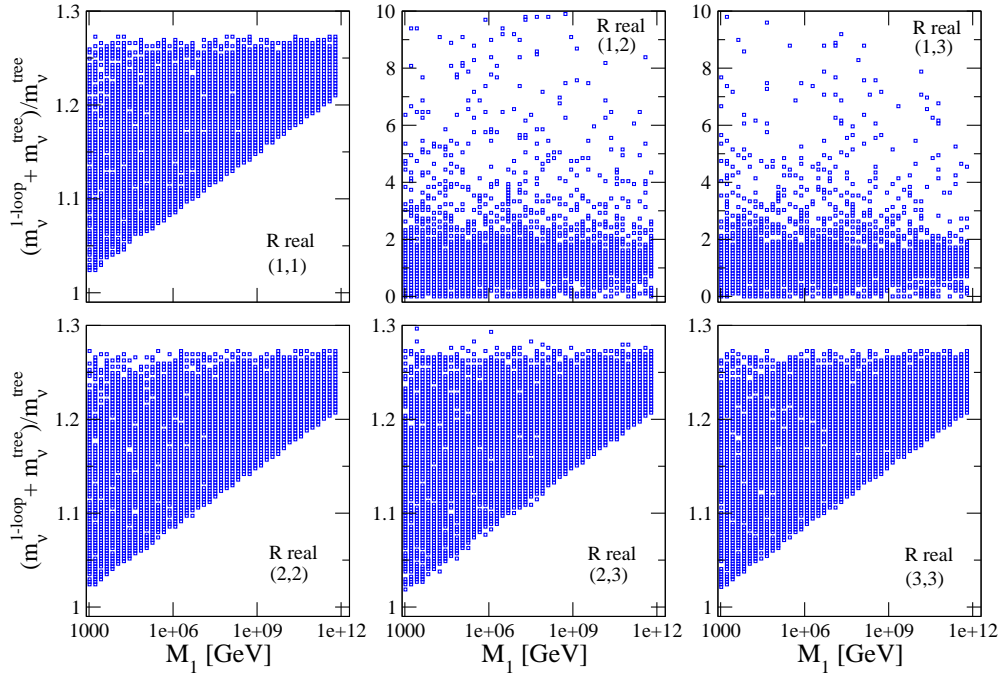}
\caption{\small \it The ratio between the 1-loop and the tree-level result for
the different elements of the neutrino mass matrix as a function of $M_1$.
It has been assumed that $\mathbf{R}$ is real and that light neutrinos have a IH
spectrum.
\label{fig:iarrayRR}}
\end{center}
\end{figure}

The finite corrections for  $\mathbf{R}$  real are shown, as a function of $M_1$,  in figure \ref{fig:iarrayRR}. The range of variation is approximately the same for 
the elements ($2,2$), ($2,3$), ($3,3$), and ($1,1$), reaching maximum values of order $30\%$ independently of $M_1$. The minimum value of the correction, on the other hand, is seen to increase with $M_1$. The entries ($1,2$) and ($1,3$) may
feature large fractional corrections, a consequence of the vanishing matrix elements at tree-level. Comparing these results with those obtained in the previous section, it is evident that the type of light neutrino spectrum does not have a decisive impact on the generic size or behavior of the finite corrections.
\subsection{$\mathbf{R}$ complex}
\begin{figure}[tb]
\begin{center}
\includegraphics[scale=0.45]{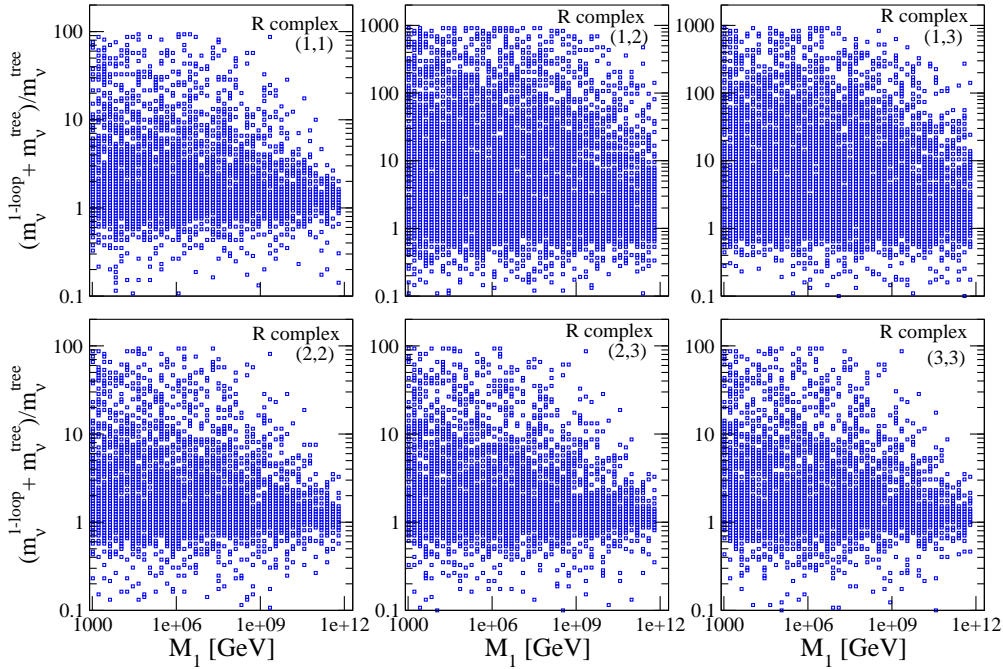}
\caption{\small \it The ratio between the 1-loop and the tree-level result for
the different elements of the neutrino mass matrix as a function of $M_1$.
It has been assumed that $\mathbf{R}$ is complex and that light neutrinos have a IH
spectrum.
\label{fig:iarrayRC}}
\end{center}
\end{figure}
In the most general case of a complex $\mathbf{R}$ matrix, the corrections tend to be quite
large for all entries, as illustrated by figure \ref{fig:iarrayRC}. They can easily reach $2$ or $3$ orders of magnitude above the tree-level value, being typically larger for the ($1,2$) and ($1,3$) matrix elements. They may also give rise to cancellations between the tree-level and the 1-loop contribution, such that the full result at 1-loop could only be a small fraction of the tree-level result --see e.g. the points around $0.1$ in the figure.

These large fractional corrections to the elements of the neutrino mass matrix translate into important deviations in the neutrino mass eigenvalues and the neutrino mixing angles, just as for the spectrum with normal hierarchy --see previous section.

For a light neutrino spectrum with inverted hierarchy, therefore, the corrections are even more important than for the normal hierarchy spectrum, as there are two different matrix elements that can receive large fractional corrections independently of $\mathbf{R}$. If $\mathbf{R}$ contains large numbers then the corrections to all matrix elements are usually significant for both the normal and the inverted hierarchy spectrum.

\section{A specific example: tribimaximal mixing}
\label{sec:tribi}
We would like now to apply the ideas discussed in the previous sections to a particular and well-motivated scenario: seesaw models with tribimaximal mixing (see e.g. \cite{Harrison:2002er}). 
In scenarios with tribimaximal mixing, the neutrino mixing matrix is given at tree level by
\begin{equation}
U_\ell=\left(\begin{array}{ccc}
\sqrt{2/3} & 1/\sqrt{3} &0\\
-1/\sqrt{6} & 1/\sqrt{3} & -1/\sqrt2\\
-1/\sqrt6 & 1/\sqrt3 & 1/\sqrt2
\end{array}\right)\,.
\end{equation}
As a result, the mixing angles take the values $\theta_{12}=35.3^\circ$, $\theta_{23}=45^\circ$, $\theta_{13}=0$ at tree level. One may therefore wonder how they would change once the finite one-loop corrections to the seesaw neutrino mass matrix that we have studied are taken into account\footnote{Additional corrections from other sources may also be important but are not considered here.}. For simplicity, we will limit ourselves in this section to the normal hierarchy spectrum and to the case $\mathbf{R}$ real. Larger corrections are expected if  $\mathbf{R}$ is complex.

\begin{figure}[tb]
\begin{center}
\includegraphics[scale=0.45]{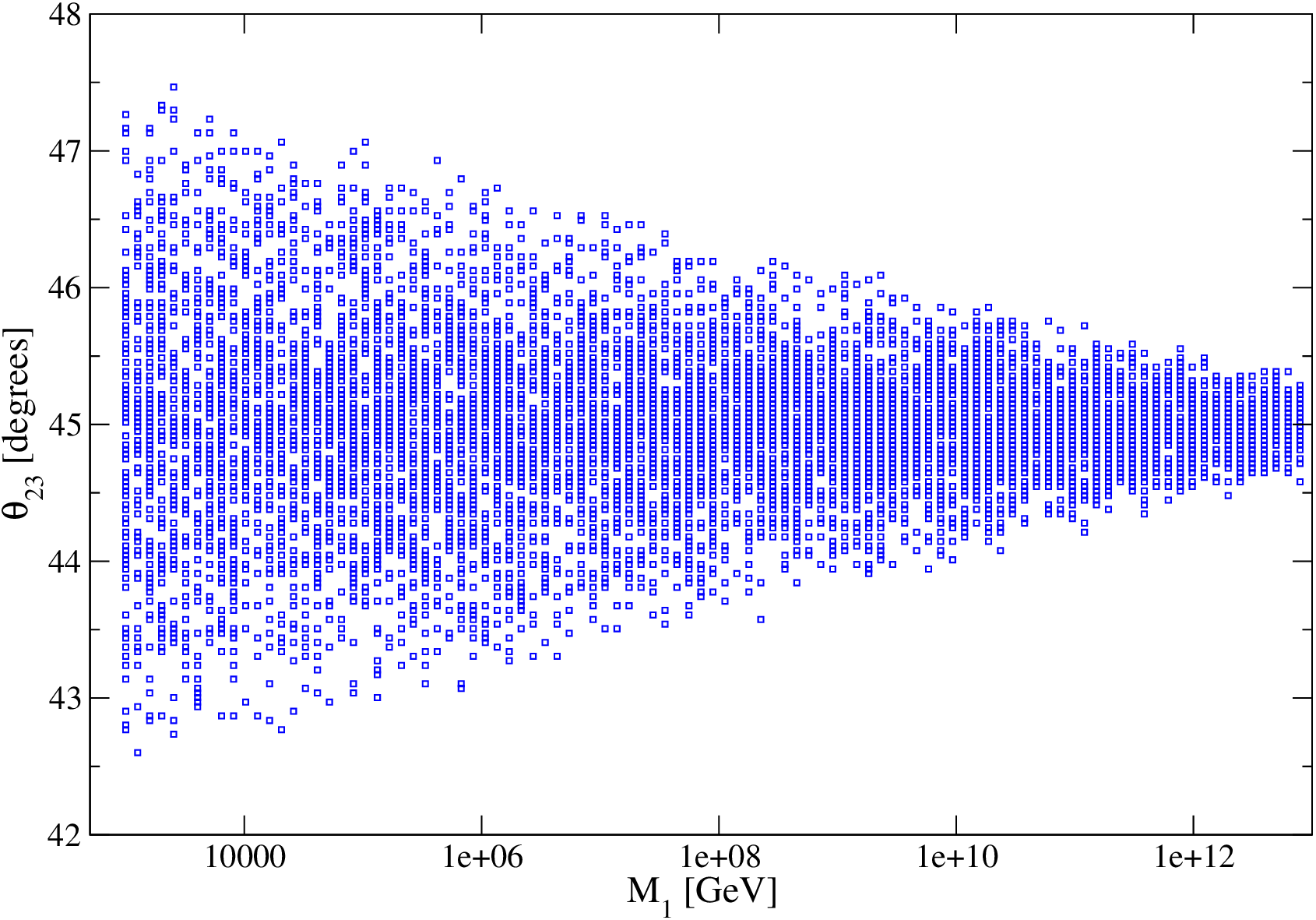}
\caption{\small \it The mixing angle $\theta_{23}$ at 1-loop as a function of the lightest heavy neutrino. At tree level $\theta_{23}$ was assumed to be $45^\circ$, in agreement with the tribimaximal mixing pattern.
\label{fig:tribit23}}
\end{center}
\end{figure}

Figure \ref{fig:tribit23} shows the 1-loop value of $\theta_{23}$ as a function of the lightest heavy neutrino mass. We see that it can deviate from its tree-level value by up to $2.5$ degrees and that the maximum deviation decreases with $M_1$. In fact, for $M_1\gtrsim 10^{12}$ GeV the correction is smaller than half a degree. This dependence with $M_1$ is very different to that observed for the matrix elements but it is consistent with it. Since the mixing angles are determined by ratios between different matrix elements, in order to have a sizable variation in the mixing angles, the structure of the neutrino mass matrix should vary significantly at 1-loop, implying that the correction should no be proportional to  the identity. Hence, in models where there is a large hierarchy between the masses of the heavy neutrinos, the corrections to the mixing angles are larger. That is exactly what is observed in figure \ref{fig:tribit23}.

\begin{figure}[tb]
\begin{center}
\includegraphics[scale=0.45]{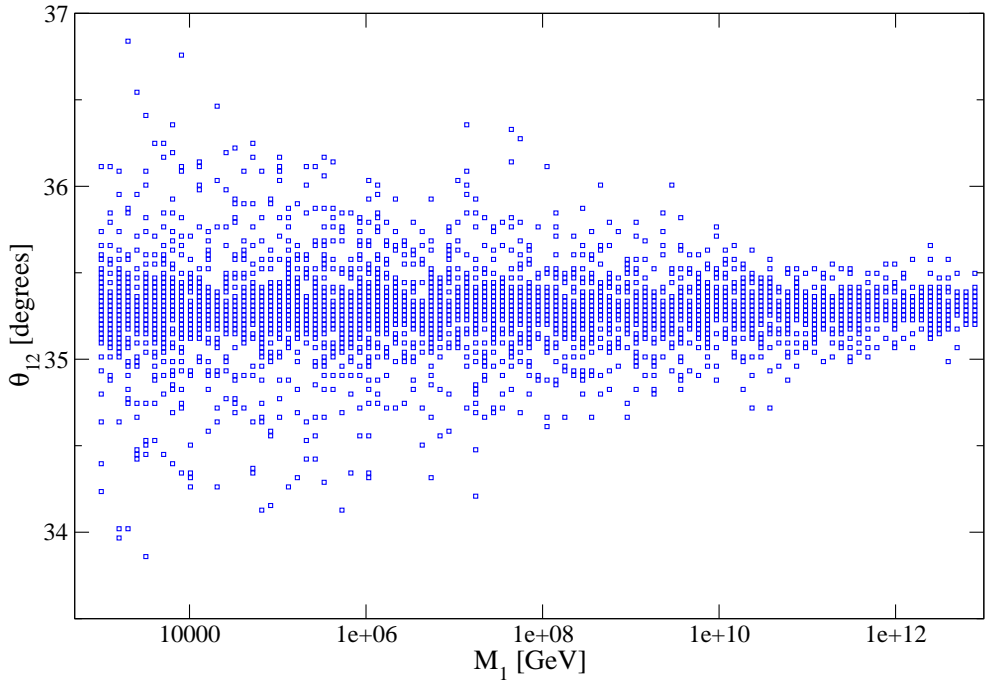}
\caption{\small \it The mixing angle $\theta_{12}$ at 1-loop as a function of the lightest heavy neutrino. At tree level $\theta_{12}$ was assumed to be $35.3^\circ$, in agreement with the tribimaximal mixing pattern.
\label{fig:tribit12}}
\end{center}
\end{figure}

\begin{figure}[tb]
\begin{center}
\includegraphics[scale=0.45]{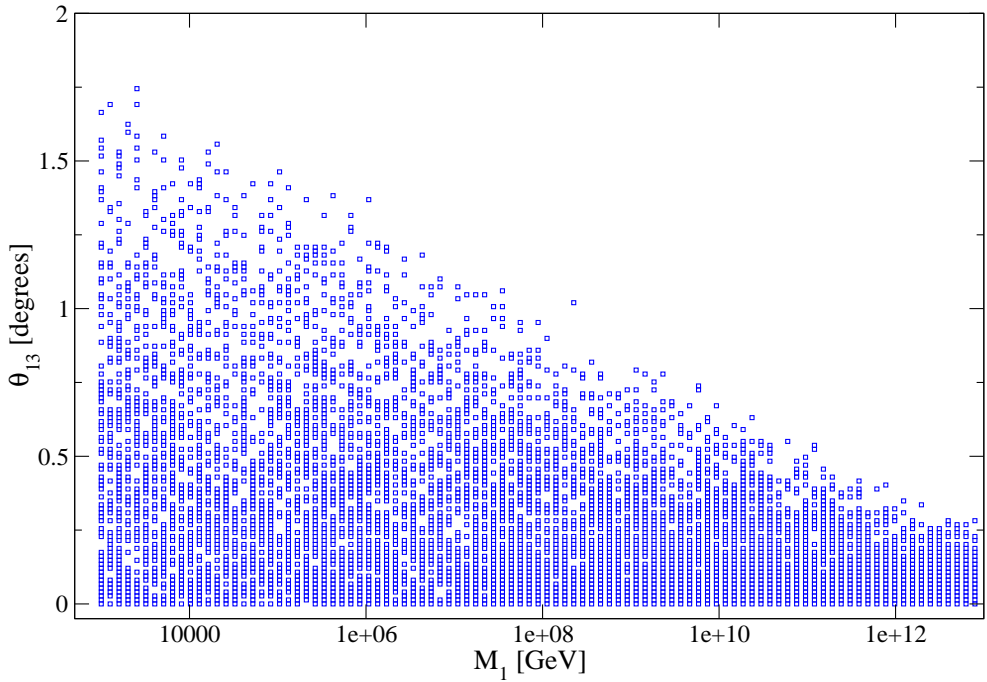}
\caption{\small \it The mixing angle $\theta_{13}$ at 1-loop as a function of the lightest heavy neutrino. At tree level $\theta_{13}$ was assumed to vanish, in agreement with the tribimaximal mixing pattern.
\label{fig:tribit13}}
\end{center}
\end{figure}

The 1-loop corrected value of $\theta_{12}$ is shown in figure \ref{fig:tribit12} as a function of $M_1$. The variation in this case is smaller and it also decreases with $M_1$. More interesting for the phenomenology of neutrinos and for  future experiments is the correction to $\theta_{13}$, which is exactly zero at tree level.  Figure \ref{fig:tribit13} shows that the 1-loop corrected value of $\theta_{13}$ can reach almost $2$ degrees, corresponding to $\sin^2\theta_{13}\sim 10^{-3}$, for $M_1$ around $1$ TeV. As expected, the maximum correction decreases with $M_1$, amounting to about $1$ degree ($\sin^2\theta_{13}\sim 3\times 10^{-4}$) for $M_1\sim 10^{8}$ GeV. Given that a neutrino factory could be sensitive to $\sin^2\theta_{13}\sim 10^{-5}$ \cite{Huber:2003ak}, these corrections are certainly within the reach of future experiments.

\section{Discussion}
\label{sec:discussion}
The huge corrections we have found for fine-tuned models  are not surprising. It is well-known in the literature that radiative corrections to fine-tuned models should spoil the tuning imposed at tree-level between the different parameters (see e.g.\cite{Ibarra:2010xw}) and that it is unnatural to expect otherwise. We have explicitly shown that that is the case. The 1-loop corrections to the neutrino mass matrix in fine-tuned seesaw models are so large that the compatibility between the tree-level seesaw formula and the experimental data becomes irrelevant. The inclusion of the 1-loop corrections is in such case necessary.

As a way of avoiding such large corrections, one may think of including the 1-loop correction into the seesaw parametrization from the very beginning. That is, one would like to look for the most general solution to equation (\ref{eq:eff-nmm-1loop}) rather than to equation (\ref{eq:seesaw-formula-tree-level}). Doing so, however, would not completely solve the problem. We would, instead, be constructing fine-tuned models at the 1-loop level, which are expected to receive large corrections at 2-loops. Fine-tuned models, it seems, are better avoided.

One possibility to do so is simply to restrict from the very beginning the magnitude of the complex angles that parametrize $\mathbf{R}$. If they are such that $\left|R_{ij}\right|_{\mbox{\tiny{max}}}\lesssim 1$ then no fine-tuning occurs and the corrections are usually under control. This additional restriction, however, has not been taken into account in previous analysis. And it was recently suggested in \cite{Casas:2010wm} that a fair scan of the seesaw parameter space is one in which no restriction beyond the perturbativity of the neutrino Yukawa couplings, which we have implemented, is imposed. Our results clearly demonstrate, on the contrary, that perturbativity is not enough to guarantee the stability of the neutrino mass matrix under radiative corrections and that wrong results can easily be reached if the 1-loop corrections are not taken into account. In fact, as we have seen, a significant fraction of models which are compatible with the data at tree level are no longer so once the 1-loop corrections are considered. Thus, when one is randomly scanning the parameter space of the seesaw model, it is necessary to include the finite corrections to neutrino masses.

Another  possible way out is the use of a different parametrization. An alternative to the $\mathbf{R}$ parametrization that has been used in previous works is  the $V_L$ parametrization. In it,  $\mathbf{M_D}$ is written as
\begin{equation}
  \label{eq:VL-param}
  \mathbf{M_D}=\mathbf{V}_{\mathbf{R}}^\dagger\,\mathbf{\hat M_D}\mathbf{V_L}\,,
\end{equation}
where $\mathbf{\hat M_D}$ is a diagonal matrix defined by the eigenvalues
of $\mathbf{M_D}$ (real and positive) and $\mathbf{V_{L,R}}$ are unitary
matrices determined by three rotation angles and three complex phases.
In this parametrization, $\mathbf{\hat M_D}$, $\mathbf{V_{L}}$ and the neutrino data are used as inputs.   Using equation (\ref{eq:VL-param}) and the effective light neutrino mass
matrix the following relation is obtained
\begin{equation}
  \label{eq:nmm-VL-param}
  \mathbf{\hat M_D}^{-1}\mathbf{V}_{\mathbf{L}}^*\mathbf{m}_{\nu}^{{\mbox{\tiny (tree)}}}
  \mathbf{V}_{\mathbf{L}}^\dagger\mathbf{\hat M_D}^{-1}=\mathbf{V_R^*}
  \mathbf{\hat M_R}^{-1}\mathbf{V_{R}^\dagger}\,,
\end{equation}
which allows us to determine $\mathbf{V_{R}}$ and $\mathbf{\hat M_R}$ for a given set of input parameters. We have computed the corrections to the neutrino mass matrix also in this parametrization, and have observed that the results for $\mathbf{R}=\mathbf{I}$ and $\mathbf{R}$ real are easily reproduced for $\mathbf{V_L}=\mathbf{I}$ and $\mathbf{V_L}$ real. In particular, the large corrections for certain matrix elements are obtained there too. An important difference occurs, however, for $\mathbf{R}$ complex. Due to the different way in which the $\mathbf{V_L}$ parametrization samples the parameter space of the seesaw model, it is way more difficult to find fine-tuned models, with the consequence that models with very large corrections are rather scarce in the $\mathbf{V_L}$ parametrization.  

In a future publication, we will discuss additional implications of these
corrections, including their evaluation in supersymmetric scenarios as well as their possible effects in leptogenesis and lepton flavor violating processes.

\section{Conclusions}
\label{sec:conclusions}
The seesaw model is one of the most appealing extensions of the Standard Model that
can explain  neutrino masses. In this model, the mass matrix of light neutrinos
receives finite corrections from 1-loop diagrams mediated by the heavy
neutrinos. We
considered the two different kinds of light neutrino spectra, hierarchical and
inverted, and computed the corrections to the entries of the neutrino mass
matrix as a function of the seesaw parameters.  We found  these
corrections to be quite important, exceeding in several cases of interest the
tree level result by orders of magnitude. Two different reasons were
identified as leading to a large correction: an unusually suppressed tree-level
result and an $\mathbf{R}$ matrix with elements much larger than $1$. Examples of
the first case are the corrections to the matrix element ($1,3$) for NH
neutrinos and to ($1,2$) and ($1,3$) for IH neutrinos. The second case can
occur for $\mathbf{R}$ complex and includes the so-called fine-tuned models considered
in the literature. Since these corrections can be large, models that at
tree-level are compatible with the experimental neutrino data will not
necessarily be so at the 1-loop level, modifying in a significant way the viable
regions in the parameter space of the seesaw model. As a particular example, we studied the corrections to the mixing angles in seesaw scenarios with tribimaximal mixing and show them to lead to observable effects in future experiments. We stressed, therefore, that because of their size and importance,
these corrections must necessarily be taken into account in the study and
analysis of  seesaw models.  

\section*{Acknowledgments}
We would like to thank Maria Jose Herrero, Enrico Nardi, Alejandro Ibarra and Walter Winter for useful discussions and suggestions. DAS is supported
by a belgian FNRS postdoctoral fellowship. CEY is supported by DFG grant no.
WI 2639/2-1.
%%%%
\appendix
\section{Finite self-energy functions}
\label{sec:self-energies}
\begin{figure}[t]
  \centering
  \includegraphics[width=7cm,height=2cm]{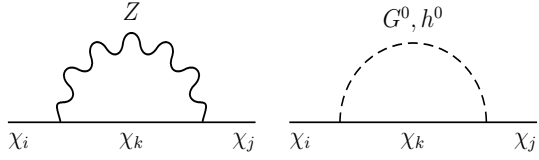}
  \caption{Self-energy diagrams accounting for $\mathbf{\delta M_L}$}
  \label{fig:slef-energy-diagrams}
\end{figure}
In this appendix we present the calculation of the finite 1-loop corrections
$\mathbf{\delta M_L}$ discussed in section \ref{sec:one-loop-corrections},
we will closely follow ref. \cite{Grimus:2002nk}.
The self-energy function $\mathbf{\Sigma_L^S}(0)$, that determines these
corrections, can be written as
\begin{equation}
  \label{eq:self-energy-deltaML}
  -i\mathbf{\Sigma_L^S}(0)=-i
  \left[
    \mathbf{\Sigma_L^S}^{(Z)}(0)+
    \mathbf{\Sigma_L^S}^{(G^0)}(0)+
    \mathbf{\Sigma_L^S}^{(h^0)}(0)
  \right]\,,
\end{equation}
where the $\mathbf{\Sigma_L^S}^{(Z,G^0,h^0)}(0)$ functions arise from the
self-energy Feynman diagrams (evaluated at zero external momentum)
involving the $Z$, the neutral Goldstone
boson $G^0$ and the Higgs $h^0$ shown in
fig. \ref{fig:slef-energy-diagrams}. The calculation of these
functions is determined by the coupling of the $Z$ with the
Majorana eigenstates $\mathbf{\chi}$:
\begin{equation}
  \label{eq:Lag-Z}
  {\cal L}_Z=
    \frac{g}{4c_w}Z_\mu\bar\chi\gamma^\mu
    \left[
      P_L(\mathbf{U_L}^\dagger\mathbf{U_L})
      -
      P_R(\mathbf{U_L}^T\mathbf{U_L}^*)
    \right]\chi\,,
\end{equation}
($c_w=\cos\theta_w$ with $\theta_w$ the weak mixing angle),
the couplings with the Higgs boson, derived from the Lagrangian
(\ref{eq:seesaw-Lag}),
\begin{equation}
  \label{neutral-L}
  -{\cal L}_{h^0}=\frac{1}{2\sqrt{2}}h^0\bar\chi
  \left[
    \mathbf{O^S_L}P_L
    +
    \mathbf{O^S_R}P_R
  \right]\chi\,,
\end{equation}
where the couplings $\mathbf{O^S_{L,R}}$ are given by
\begin{align}
  \label{eq:couplings}
  \mathbf{O^S_L}=&\mathbf{U}_{\mathbf{R}}^\dagger\mathbf{\lambda}\mathbf{U}_{\mathbf{L}}
  + \mathbf{U}_{\mathbf{L}}^T\mathbf{\lambda}^T\mathbf{U}_{\mathbf{R}}^*\\
  \mathbf{O^S_R}=&\mathbf{U}_{\mathbf{L}}^\dagger\mathbf{\lambda}^\dagger\mathbf{U_R}
  + \mathbf{U}_{\mathbf{R}}^T\mathbf{\lambda}^*\mathbf{U}_{\mathbf{L}}^*\,,
\end{align}
and finally the couplings with $G^0$ that can be obtained from the
Lagrangian (\ref{neutral-L}) by replacing $\mathbf{O^S_{L,R}}\to
-i\mathbf{O^S_{L,R}}$.
From these Lagrangians and the diagonalization
relation~(\ref{eq:loop-parameters}) it can be seen that
$\mathbf{\Sigma_L^S}^{(Z)}(0)$ contributes only to $\mathbf{\delta
  M_L}$ whereas $\mathbf{\Sigma_L^S}^{(G^0,h^0)}(0)$ contribute to all
the block matrices of the $6\times 6$ neutral fermion mass matrix.
The contributions of these self-energies to $\mathbf{\delta M_L}$
can be identified by means of the relation
(\ref{eq:finite-contribution}).

Using dimensional regularization ($d=4-\epsilon$) and working in the
$R_\xi$ gauge the $Z$ self-energy function is found to be
\begin{equation}
  \label{eq:self-energy-Z}
  \mathbf{\Sigma_L^S}^{(Z)}(0)=
  \mathbf{U_L}^T
  \left[
    \mathbf{\delta M_L}^{(Z)}(1)
    +
    \mathbf{\delta M_L}^{(Z)}(2,1)
    +
    \mathbf{\delta M_L}^{(Z)}(2,2)
  \right]
  \mathbf{U_L}\,,
\end{equation}
where the different matrices can be expressed in terms of the
Passarino-Veltman function $B_0(0,m_0^2,m_1^2)$ \cite{Passarino:1978jh},
namely
\begin{equation}
  \label{eq:deltamZ1}
  \mathbf{\delta M_L}^{(Z)}(1)=-\frac{g^2}{64\pi^2c^2_w}(4-\epsilon)
  \mathbf{U_L}^*\;\mathbf{\hat {\cal M}}\;B_0(0,M_Z^2,\mathbf{\hat {\cal M}}^2)\;
  \mathbf{U_L}^\dagger\,,
\end{equation}
\begin{equation}
  \label{eq:deltamZ21}
  \mathbf{\delta M_L}^{(Z)}(2,1)=-\frac{g^2}{64\pi^2c^2_w M_Z^2}
  \mathbf{U_L}^*\;\mathbf{\hat {\cal M}}^3\;B_0(0,\xi_Z M_Z^2,\mathbf{\hat {\cal M}}^2)\;
  \mathbf{U_L}^\dagger\,,
\end{equation}
\begin{equation}
  \label{eq:deltamZ22}
  \mathbf{\delta M_L}^{(Z)}(2,2)=\frac{g^2}{64\pi^2c^2_w M_Z^2}
  \mathbf{U_L}^*\;\mathbf{\hat {\cal M}}^3\;B_0(0,M_Z^2,\mathbf{\hat {\cal M}}^2)\;
  \mathbf{U_L}^\dagger\,.
\end{equation}
As regards the $G^0$ and $h^0$ self-energies they are given by
\begin{equation}
  \label{eq:self-energy-G0h0}
  \mathbf{\Sigma_L^S}^{(X)}=\mathbf{U_L}^T\mathbf{\delta M_L}^{(X)}
  \mathbf{U_L}\qquad (X=G^0, h^0)
\end{equation}
 with $\mathbf{\delta M_L}^{(X)}$ given by
\begin{align}
  \label{eq:deltaMLG0}
  \mathbf{\delta M_L}^{(G^0)}&=\frac{g^2}{64\pi^2c^2_w M_Z^2}
  \mathbf{U_L}^*\;\mathbf{\hat {\cal M}}^3\;B_0(0,\xi_Z M_Z^2,\mathbf{\hat {\cal M}}^2)\;
  \mathbf{U_L}^\dagger\\
  \label{eq:deltaMLh0}
  \mathbf{\delta M_L}^{(h^0)}&=-\frac{g^2}{64\pi^2c^2_w M_Z^2}
  \mathbf{U_L}^*\;\mathbf{\hat {\cal M}}^3\;B_0(0,m_h^2,\mathbf{\hat {\cal M}}^2)\;
  \mathbf{U_L}^\dagger\,.
\end{align}
In the calculation of the above expressions we have used
the relation
\begin{equation}
  \label{eq:unitarity-relations}
  \mathbf{U_R}^\dagger\;\mathbf{M_D}=\mathbf{\hat {\cal M}}\;
  \mathbf{U_L}^\dagger\,,
\end{equation}
that follows from the diagonalization relation
(\ref{eq:diag-matrix}) and the unitarity constraints of the matrix $\mathbf{U}$.

Some words are in order regarding these results.  Corrections
(\ref{eq:deltamZ21}) and (\ref{eq:deltaMLG0}) cancel, ensuring the
gauge invariance of the result. The Passarino-Veltman function $B_0$
has a finite and infinite part\footnote{Here by infinite part we mean
$B_0^{(inf)}(0,m_0^2,m_1^2)=2\epsilon^{-1}-\gamma+4\pi+1$.},
the infinite piece in
$\mathbf{\delta M_L}^{(Z)}(1)$ cancels due to the constraint
\begin{equation}
  \label{eq:unitarity-constraint}
  \mathbf{U_L}^*\;\mathbf{\hat {\cal M}}\;\mathbf{U_L}=\mathbf{0}\,,
\end{equation}
whereas the divergent pieces in $\mathbf{\delta M_L}^{(Z)}(2,2)$ and
$\mathbf{\delta M_L}^{(h^0)}$ cancel among them\footnote{There is also
  a finite term, $\ln \mathbf{\hat {\cal M}}^2$, that cancels.}, thus demonstrating
that $\mathbf{\delta M_L}$ is finite as anticipated in
sec. \ref{sec:one-loop-corrections}. Taking into account that the
finite part of $B_0$ can be recasted as
\begin{align}
  \label{eq:passarino-veltman-fin}
  B_0^f(0,m_0^2,m_1^2)&=
  -\left[
    \frac{1}{m_1^2/m_0^2-1}\ln \left(\frac{m_1^2}{m_0^2}\right)
    +
    \ln m_1^2
  \right]\nonumber\\
  &=-\left[\frac{m_1^2/m_0^2\; \ln \left(\frac{m_1^2}{m_0^2}\right)}
    {m_1^2/m_0^2-1} + \ln m_0^2\right]
\end{align}
the finite parts of $\mathbf{\delta M_L}^{(Z)}(1)$ and
$\mathbf{\delta M_L}^{(Z)}(2,2)$ combine to yield
\begin{equation}
  \label{eq:finite-deltaMLZ}
  \mathbf{\delta M_L}^{(Z)f}=\frac{3g^2}{64\pi^2M_W^2}
  \mathbf{U_L}^*\mathbf{\hat {\cal M}}^3
  \left(
    \frac{\mathbf{\hat {\cal M}}^2}{M_Z^2}-\mathbf{1}
  \right)^{-1}
  \ln\left(\frac{\mathbf{\hat {\cal M}}^2}{M_Z^2}\right)
  \mathbf{U_L}^\dagger\,.
\end{equation}
Finally the finite contribution from the Higgs self-energy function
reads
\begin{equation}
  \label{eq:deltaML-H0fin}
   \mathbf{\delta M_L}^{(h^0)f}=\frac{g^2}{64\pi^2M_W^2}
  \mathbf{U_L}^*\mathbf{\hat {\cal M}}^3
  \left(
    \frac{\mathbf{\hat {\cal M}}^2}{m_{h^0}^2}-\mathbf{1}
  \right)^{-1}
  \ln\left(\frac{\mathbf{\hat {\cal M}}^2}{m_{h^0}^2}\right)
  \mathbf{U_L}^\dagger\,.
\end{equation}

The finite correction $\mathbf{\delta M_L}$, discussed in sec.
\ref{sec:one-loop-corrections}, is obtained from the dominant parts of
eqs. (\ref{eq:finite-deltaMLZ}) and (\ref{eq:deltaML-H0fin}) (order $\mathbf{\hat M_R}^{-1}$).
These pieces can be extracted by using
eq. (\ref{eq:unitarity-relations}) and by taking into account
that in the seesaw limit $\mathbf{M_D}\ll \mathbf{M_R}$, in the basis for
which $\mathbf{M_R}$ is diagonal, the matrix $\mathbf{U_R}$ can be
written as
%\begin{equation}
%  \label{eq:U-approx}
%  \mathbf{U}=
%  \begin{pmatrix}
%    \mathbf{1} & \mathbf{\xi}\\
%    -\mathbf{\xi}^\dagger & \mathbf{1}
%  \end{pmatrix}
%  \begin{pmatrix}
%    \mathbf{U_\ell} & \mathbf{0}\\
%    \mathbf{0} & \mathbf{1}
%  \end{pmatrix}\quad \mbox{with}\quad\mathbf{\xi}=\mathbf{M_D}^T\,
%  \mathbf{M_R^{-1}}\,.
%\end{equation}
%The first matrix above block-diagonalize the neutral fermion mass matrix
%approximately to the form $\mbox{diag}(\mathbf{m}_{\nu}^{\mbox{\tiny(tree)}}, \mathbf{\hat M_R})$.
%Accordingly, the matrices $\mathbf{U_{L,R}}$ become
\begin{align}
 % \label{eq:ULapprox}
 % \mathbf{U_L}&=(\mathbf{U_\ell},\mathbf{\xi})\,,\\
  \label{eq:URapprox}
  \mathbf{U_R}&=(-\mathbf{\xi}^\dagger\mathbf{U_\ell},\mathbf{1})\,,
\end{align}
where $\mathbf{\xi}=\mathbf{M_D}^T\,\mathbf{M_R^{-1}}$.
%By doing so, the result (\ref{eq:deltaML}) is obtained.
%presented in sec. \ref{sec:one-loop-corrections}
%From (\ref{eq:unitarity-relations}) and using (\ref{eq:URapprox}) the dominant
%contributions to $\mathbf{\delta M_L}$ (order $\mathbf{\hat M_R}^{-1}$) can be extracted from
%eqs. (\ref{eq:finite-deltaMLZ}) and (\ref{eq:deltaML-H0fin}). The result
%is given by expression (\ref{eq:deltaML}) in sec. \ref{sec:one-loop-corrections}.

\bibliographystyle{unsrt}
\bibliography{neutrino}

\end{document}